\documentclass[prl,floatfix,twocolumn,showpacs,preprintnumbers,amsmath,amssymb,superscriptaddress]{revtex4}
\usepackage{graphicx,float,datetime}
\usepackage[ansinew]{inputenc}

\newcommand{\udt}[3]{#1^{#2}_{\phantom{#2}#3}}

\newdateformat{mydate}{\THEDAY\hspace{3pt}\monthname[\THEMONTH] \THEYEAR}

\begin{document}

\begin{center}
\title{Charged Cylindrical Black Holes in Conformal Gravity}
\date{\mydate\today}
\author{Jackson Levi Said\footnote{jsai0004@um.edu.mt}}
\affiliation{Physics Department, University of Malta, Msida, MSD 2080, Malta}
\author{Joseph Sultana\footnote{joseph.sultana@um.edu.mt}}
\affiliation{Mathematics Department, University of Malta, Msida, MSD
2080, Malta}
\author{Kristian Zarb Adami\footnote{kristian.zarb-adami@um.edu.mt}}
\affiliation{Physics Department, University of Malta, Msida, MSD 2080, Malta}
\affiliation{Physics Department, University of Oxford, Oxford, OX1 3RH, United Kingdom}

\begin{abstract}
{Considering cylindrical topology, we present the static solution for
a charged black hole in conformal gravity. We show that
unlike the general relativistic case, there are two different
solutions, both including a factor which gives rise to a linear term
in the potential, which also features in the neutral case. This may
have significant ramifications for particle trajectories.}
\end{abstract}


\maketitle

\end{center}

\section{I. Introduction}
Einstein's theory of general relativity has succeeded
extraordinarily well in solar system observations. However, when
larger length scales are investigated, an overwhelming amount of dark
mass energy must be introduced in order to reproduce observations
such as with galactic rotation curves and the accelerating expansion
of the Universe. It may be true that most of the contributing
mass-energy of the Universe is nonluminous, but it may also be the
case that the underlying theory contains other factors whose
contribution only becomes significant on large and very large
scales.
\newline

A number of proposed models have attempted to add terms and factors
which only become significant on the large scales such as with MOND
\cite{0p03} and $f\left(R\right)$ \cite{p04} gravity. On the other
hand, other proposals aim to exploit some hidden assumption or
principle in general relativity. One such idea is the fourth-order
conformal Weyl gravity model introduced in
Refs.\cite{riegert,conformal2} which is based on the underlying
principle of local conformal invariance such that the manifold
remains the same under local conformal stretchings of the kind
\begin{equation}
g_{\mu\nu}\left(x\right)\rightarrow\Omega^2\left(x\right)g_{\mu\nu}\left(x\right).
\label{metr_conf_trans}
\end{equation}
This restrictive invariance principle leads to a fourth-order theory
with the unique action \cite{conformal2}
\begin{align}
I_W&=\displaystyle\int d^4x\sqrt{-g} L\nonumber\\
&=-\alpha_c\displaystyle\int d^4x\sqrt{-g}C_{\lambda\mu\nu\kappa}C^{\lambda\mu\nu\kappa}\nonumber\\
&=-2\alpha_c\displaystyle\int d^4x\sqrt{-g}\left[R_{\mu\nu}R^{\mu\nu}-\frac{1}{3}R^2\right],
\label{weyl_action}
\end{align}
where $\alpha_c$ is a dimensionless coupling constant and the Weyl tensor $C_{\lambda\mu\nu\kappa}$ is given by
\begin{align}
C_{\lambda\mu\nu\kappa}&=R_{\lambda\mu\nu\kappa}\nonumber\\
&-\frac{1}{2}\left(g_{\lambda\nu}R_{\mu\kappa}-g_{\lambda\kappa}R_{\mu\nu}-g_{\mu\nu}R_{\lambda\kappa}+g_{\mu\kappa}R_{\lambda\nu}\right)\nonumber\\
&+\frac{1}{6}R\left(g_{\lambda\nu}g_{\mu\kappa}-g_{\lambda\kappa}g_{\mu\nu}\right).
\end{align}
This tensor also satisfies the conformal principle
\begin{equation}
C_{\lambda\mu\nu\kappa}\rightarrow\tilde{C}_{\lambda\mu\nu\kappa}=\Omega^2\left(x\right)\;C_{\lambda\mu\nu\kappa},
\end{equation}
due to its dependence on the metric tensor.
\newline

An immediate consequence of taking this action is that the
cosmological length scale, which appears in general relativity through
the cosmological constant $\Lambda$, is unnecessary here. One of the
interesting consequences of conformal gravity is that a number of
behaviors still emerges despite not considering a cosmological
constant such as the fact that the Schwarzschild-de Sitter metric
\cite{conformal2} is also a solution to the field equations of Weyl
gravity. Besides this, it has been shown \cite{furthersolutions4}
that conformal gravity despite being a fourth-order theory, still
admits a Newtonian potential $1/r$ term in the field of any
spherically symmetric matter distribution described by a fourth-order Poisson equation. Therefore, although the second-order Poisson
equation in general relativity is sufficient to generate a Newtonian
potential, it is not by any means a necessary requirement, so that
Newton's law of gravity remains valid in the fourth-order Weyl
gravity.
\newline
The conformal action of Eq.(\ref{weyl_action}) leads to the field
equations \cite{p02}
\begin{align}
(-g)^{-1/2}g_{\mu\alpha}g_{\nu\beta}\frac{\delta I_{W}}{\delta
g_{\alpha\beta}}
&=-2\alpha_{c} W_{\mu\nu}\nonumber\\
&=-\frac{1}{2}T_{\mu\nu},
\label{weyl_field_eqns}
\end{align}
where $T_{\mu\nu}$ is the stress-energy tensor, and
\begin{equation}
W_{\mu\nu} = 2C^{\alpha\ \ \beta}_{\ \mu\nu\ ;\beta\alpha} +
C^{\alpha\ \ \beta}_{\ \mu\nu}R_{\alpha\beta}, \label{w1}
\end{equation}
is the Bach tensor. Thus, when the Ricci tensor $R_{\mu\nu}$ vanishes,
so does $W_{\mu\nu}$, implying that any vacuum solution of general
relativity also carries over to conformal gravity naturally. However
the converse does not hold in general since there are other ways by
which the Bach tensor can vanish.
\newline

At present, there are a number of solutions to the conformal gravity
field equations. These include the static, spherically symmetric
vacuum case obtained by Mannheim and Kazanas in Ref.
\cite{conformal2},
\begin{equation}
ds^2 = -B(r)\,dt^2 + \frac{dr^2}{B(r)} + r^2(d\theta^2 +
\sin^2\theta\,d\phi^2),
\end{equation}
where
\begin{equation}
B(r) = 1  - \frac{\beta(2 - 3\beta\gamma)}{r} - 3\beta\gamma +
\gamma r - kr^2, \label{mkmetric}
\end{equation}
which includes the Schwarzschild $(\gamma = k = 0)$ and
Schwarzschild-de Sitter $(\gamma = 0)$ solutions as special cases,
and also its charged generalization \cite{riegert,furthersolutions1}
with
\begin{align}
B(r)&=(1 - 3\beta\gamma) - \frac{(\beta(2 - 3\beta\gamma) +
Q^2/8\alpha_{c}\gamma)}{r}\nonumber\\
&+ \gamma r - k r^2, \label{mkcmetric}
\end{align}
where $Q$ is the charge. Some interesting work has also been
done in Refs.\cite{klemm, Cognola, Pang} where general topological
solutions in Weyl gravity were investigated. The particular case of
cylindrically symmetric solutions in Weyl gravity has been
considered in Refs.\cite{p06,p07}, but due to difficulties in the
complex field equations, a particular gauge is chosen that does not
naturally generalize the well known cylindrically symmetric
solutions in general relativity. So in Ref.\cite{0p01}, we derived
analytically the metric for a neutral static cylindrical spacetime
by adopting a gauge similar to that used by Mannheim and Kazanas for
the spherically symmetric case to obtain
\begin{equation}
ds^2=-A\left(r\right)dt^2+B\left(r\right)dr^2+r^2d\phi^2+\alpha^2r^2dz^2,
\end{equation}
where
\begin{equation}
A^{-1}\left(r\right)=B\left(r\right)=\frac{\beta}{r}+\sqrt{\frac{3\beta\gamma}{4}}+\frac{\gamma
r}{4}+k^2r^2. \label{neut_metric}
\end{equation}
This generalizes the static black string solution in general
relativity represented by the Lemos metric \cite{0p02}
\begin{align}
ds^2&=-\left(\alpha^2r^2-\frac{b}{\alpha r}\right)dt^2+\frac{dr^2}{\alpha^2r^2-\frac{b}{\alpha r}}\nonumber\\
&+r^2\,d\phi^2 + \alpha^2r^2\,dz^2, \label{lemos_metr}
\end{align}
with $k = \alpha = \sqrt{-\Lambda/3}$ and $\beta = -b/\alpha =
-4GM/\alpha$, so that in our case, we have a linear $\gamma r$ term
in the metric similar to the spherically symmetric case in
Eq.(\ref{mkmetric}). We now seek to consider the charged case and find
that the field equations which follow by taking a nonzero stress-energy tensor in Ref.\cite{0p01} have two separate solutions, each
representing a conformal generalization of the charged black string
solution in Einstein's second-order theory \cite{0p02},
\begin{align}
ds^2&=-\left(\alpha^2r^2-\frac{b}{\alpha r}+\frac{c^2}{\alpha^2 r^2}\right)dt^2\nonumber\\
&+\frac{dr^2}{\alpha^2r^2-\frac{b}{\alpha r}+\frac{c^2}{\alpha^2
r^2}}+r^2d\phi^2+\alpha^2dz^2, \label{lem_rel}
\end{align}
where $\alpha$ and $\beta$ are the same constants as in
Eq.(\ref{lemos_metr}), and $c^2 = 4G\lambda^2$, where $\lambda$ is the
linear charge density along the $z$ axis.
\newline
The outline of the paper is as follows. In Sec. II,  we derive the
metric for the charged cylindrically symmetric spacetime in
conformal gravity and compare it with the general relativity
analogue in Eq.(\ref{lem_rel}). In Sec. III, we discuss some of
the thermodynamical properties of this new solution and then present
our discussion and conclusion in Sec. IV. The signature used in
this paper is $\left(-,+,+,+\right)$ and units where $G=1=c$ are
used.

\section{II. The Conformal Cylindrical Metric}
Consider first a general line element in cylindrical coordinates
$\left(t,\,\rho,\,\phi,\,z\right)$
\begin{equation}
ds^2=-a\left(\rho\right)\,dt^2+b\left(\rho\right)\,d\rho^2+c\left(\rho\right)\,d\phi^2+d\left(\rho\right)\,dz^2.
\label{gen_lin}
\end{equation}
The metric elements are taken to depend only on the radial
coordinate since a static cylindrically symmetric background metric
is not expected to be curved in the angular and axial directions.
\newline

Given the local conformal invariant symmetry, the metric in
Eq.(\ref{gen_lin}) can be reformulated similarly to
Refs.\cite{conformal2, furthersolutions1}; that is, given an
arbitrary function of a spacelike coordinate parameter $r$,
$\rho\left(r\right)$, the metric can be written as
\begin{align}
ds^2&=\rho^2\left(r\right)\big[-\frac{a\left(\rho\right)}{\rho^2\left(r\right)}\,dt^2+\frac{b\left(\rho\right)}{\rho^2\left(r\right)}\,d\rho^2\nonumber\\
&+\frac{c\left(\rho\right)}{\rho^2\left(r\right)}\,d\phi^2+\frac{d\left(\rho\right)}{\rho^2\left(r\right)}\,dz^2\big].
\end{align}

Now a choice can be made on this dependence constrained by the aim
of having an end result metric which is computationally less
intensive to solve for element components. We take
\begin{equation}
\displaystyle\int\frac{d\rho}{\rho^2\left(r\right)}=-\frac{1}{\rho\left(r\right)}=\displaystyle\int\frac{dr}{\sqrt{a\left(r\right)b\left(r\right)}},
\end{equation}
which then yields the metric
\begin{align}
ds^2&=\rho^2\bigg[-A\left(r\right)\,dt^2+\frac{dr^2}{A\left(r\right)}+C\left(r\right)\,d\phi^2\nonumber\\
&+D\left(r\right)\,dz^2\bigg],
\end{align}
where $A\left(r\right)=\frac{a\left(r\right)}{\rho^2\left(r\right)}$, $C\left(r\right)=\frac{c\left(r\right)}{\rho^2\left(r\right)}$, and $D\left(r\right)=\frac{d\left(r\right)}{\rho^2\left(r\right)}$.
\newline

The metric is thus conformally related to the standard cylindrical
metric for static spacetimes. Following Eq.(\ref{metr_conf_trans})
we take a transformation,
\begin{equation}
g_{\mu\nu}\rightarrow\rho^{-2}\left(r\right)g_{\mu\nu},
\end{equation}
which molds the metric into
\begin{equation}
ds^2=-A\left(r\right)\,dt^2+\frac{dr^2}{A\left(r\right)}+C\left(r\right)\,d\phi^2+D\left(r\right)\,dz^2,
\label{con_cyl_met}
\end{equation}
which is indeed more representational of the actual degrees of
freedom enjoyed by the metric. Hence, the metric will be resolved up
to an overall $r$-dependent conformal factor.
\newline

Through Eq.(\ref{weyl_field_eqns}), it follows that $W^{\mu\nu}$ can
be expressed in terms of the conformally invariant stress-energy
tensor through
\begin{equation}
W^{\mu\nu}=\frac{1}{4\alpha_{c}}T^{\mu\nu}, \label{W-T}
\end{equation}
which implies that the mass-energy information about the system is
completely contained in $W^{\mu\nu}$. Furthermore, given the
stress-energy tensor of the system, it is directly proportional to
$W^{\mu\nu}$.
\newline

Now, to actually calculate the $W^{\mu\nu}$ tensor, we first note that
in order to achieve a conformal generalization of the charged
cylindrical metric  in the Lemos gauge, we take a vector potential
\cite{0p02}
\begin{equation}
A_{\mu}=\left(-\frac{2\lambda}{\alpha r}+\text{const},0,0,0\right),
\end{equation}
where $\lambda$ and $\alpha$ have the same meaning as in the Lemos
metric in Eq.(\ref{lem_rel}). This vector potential leads to the only
nonvanishing electric field component
$E_r=F_{10}=\frac{2\lambda}{\alpha r^2}$, which is covariantly
conserved in general and does not depend on the conformal nature of
the model being considered since Maxwell's theory is conformally
invariant. The Maxwell stress-energy tensor is found to only have
one nonvanishing component,
\begin{equation}
\udt{T}{r}{r}=-\frac{2\lambda^2}{\alpha^2 r^4},
\label{r-r_se}
\end{equation}
which surprisingly is independent of metric components in
Eq.(\ref{con_cyl_met}). Besides taking cylindrical coordinates, we
also take
\begin{equation}
C\left(r\right)=r^2,
\end{equation}
since the background spacetime has cylindrical topology. Furthermore,
since we are attempting to find a conformal generalization of the
charged Lemos metric, we take the Lemos gauge
\begin{equation}
D\left(r\right)=\alpha^2r^2,
\end{equation}
in the spirit of Ref.\cite{conformal2}.
\newline

In general relativity, the common method of finding solutions is to
compare the curvature part of the field equations with the simplest
representation of the stress-energy components. However in conformal
gravity, this is far too difficult for the nontrivial instances given
that the curvature part is replaced by the Bach tensor in
Eq.(\ref{w1}). The alternative method used in Ref.\cite{conformal2}
is to consider the Euler-Lagrange equations using the general line
element
\begin{equation}
ds^2=-B\left(r\right)\,dt^2+A\left(r\right)\,dr^2+r^2d\,\phi^2+\alpha^2r^2\,dz^2,
\label{metr_drv_init}
\end{equation}
where the Lemos gauge is considered, and the coefficients of $dr^2$
and $dt^2$ are left unrelated at first.
\newline

The Euler-Lagrange equations turn out to be second order \cite{conformal2}
\begin{align}
&\sqrt{-g}\,W^{\mu\mu}=\frac{\delta I}{\delta g_{\mu\mu}}=\frac{\partial}{\partial g_{\mu\mu}}\left(\sqrt{-g}\tilde{L}\right)\nonumber\\
&-\frac{\partial}{\partial x^{\mu}}\left(\sqrt{-g}\frac{\partial
\tilde{L}}{\partial
\left(g_{\mu\mu}\right)'}\right)+\frac{\partial^2}{\partial
\left(x^{\mu}\right)^2}\left(\sqrt{-g}\frac{\partial
\tilde{L}}{\partial \left(g_{\mu\mu}\right)''}\right), \label{el}
\end{align}
where ' indicates differentiation with respect to $r$ and $\tilde{L}
= R_{\mu\nu}R^{\mu\nu} - R^2/3$.

Taking the variation with respect to $A\left(r\right)$ and
$B\left(r\right)$ yields
\begin{widetext}
\begin{align}
\sqrt{\alpha^2r^4AB}\,W^{rr}&= -\frac{\alpha ^2}{48A(r)^4 B(r)^3\sqrt{r^4\alpha^2A\left(r\right)B\left(r\right)}}\bigg[-7 r^2 B(r)^2 A'(r)^2 \left(r B'(r)-2 B(r)\right)^2\nonumber\\
&+2 r^2 A(r) B(r) \left(2 B(r)-r B'(r)\right) \big(4B(r)^2 A''(r)+3 r A'(r) B'(r)^2\nonumber\\
&-2 B(r) \left(r A''(r) B'(r)+A'(r) \left(2 r B''(r)+B'(r)\right)\right)\big)\nonumber\\
&+A(r)^2\bigg(-7 r^4 B'(r)^4+4 r^3 B(r) B'(r)^2 \left(3 r B''(r)+5 B'(r)\right)\nonumber\\
&+4 r^2 B(r)^2 \left(r^2 B''(r)^2+B'(r)^2-2 r B'(r)\left(r B^{(3)}(r)+6 B''(r)\right)\right)\nonumber\\
&+16 r B(r)^3 \left(r \left(r B^{(3)}(r)+2 B''(r)\right)-2 B'(r)\right)+16B(r)^4\bigg)\bigg],
\label{wrr}
\end{align}
\end{widetext}
and
\begin{widetext}
\begin{align}
\sqrt{\alpha^2r^4AB}\,W^{tt}&=\frac{-\alpha ^2}{48A(r)^4B(r)^4\sqrt{r^4\alpha^2A(r)B(r)}}\bigg[ 56 r^3 B(r)^3 A'(r)^3 \left(r B'(r)-2 B(r)\right)\nonumber\\
&+r^2 A(r) B(r)^2 A'(r) \bigg(57 r^2 A'(r) B'(r)^2-4 r B(r)\bigg(13 r A''(r) B'(r)\nonumber\\
&+A'(r) \left(19 r B''(r)+13 B'(r)\right)\bigg)+4 B(r)^2 \left(26 r A''(r)+7 A'(r)\right)\bigg)\nonumber\\
&+2r A(r)^2 B(r) \bigg(29 r^3 A'(r) B'(r)^3-6 r^2 B(r) B'(r) \big(2 r A''(r) B'(r)\nonumber\\
&+A'(r) \left(9 r B''(r)+4B'(r)\right)\big)+4 r B(r)^2 \bigg(A'(r) \left(r \left(6 r B^{(3)}(r)+13 B''(r)\right)-5 B'(r)\right)\nonumber\\
&+r \left(4 r A''(r)B''(r)+\left(r A^{(3)}(r)+3 A''(r)\right) B'(r)\right)\bigg)\nonumber\\
&+8 B(r)^3 \left(2 A'(r)-r \left(rA^{(3)}(r)+A''(r)\right)\right)\bigg)+A(r)^3 \bigg(-16 r^3 \left(4 B^{(3)}(r)+r B^{(4)}(r)\right) B(r)^3\nonumber\\
&+49 r^4 B'(r)^4-4r^3 B(r) B'(r)^2 \left(29 r B''(r)+11 B'(r)\right)\nonumber\\
&+4 r^2 B(r)^2 \left(9 r^2 B''(r)^2-5 B'(r)^2+2 r B'(r) \left(6 rB^{(3)}(r)+13 B''(r)\right)\right)+16 B(r)^4\bigg)\bigg].
   \label{wtt}
\end{align}
\end{widetext}

The other two elements $W^{\phi\phi}$ and $W^{zz}$ do not need to be
taken into account since we have a sufficient number of constraints;
these two further equations provide us with an independent check of
any solution which results.
\newline

Taking the time-time and radial-radial components of the metric to
be the reciprocal of each other,
$A\left(r\right)=1/B\left(r\right)$, turns Eq.(\ref{wrr}) into
\begin{align}
12 r^4\frac{1}{B}W^{rr}&=-4B^2-\nonumber\\
&4rB\left(-2B'+r\left(B''+rB'''\right)\right)+\nonumber\\
&r^2\left(-4B'^2-r^2B''^2+2rB'\left(2B''+rB'''\right)\right).
\end{align}
But by Eq.(\ref{W-T}),
\begin{equation}
W^{rr}=-\frac{\lambda^2}{2\alpha_c\alpha^2r^4}B.
\end{equation}

This eliminates both metric tensor components and radial coordinate
factors from the constraint so that
\begin{align}
-\frac{6\lambda^2}{\alpha_c\alpha^2}&=-4B^2-\nonumber\\
&4rB\left(-2B'+r\left(B''+rB'''\right)\right)+\nonumber\\
&r^2\left(-4B'^2-r^2B''^2+2rB'\left(2B''+rB'''\right)\right).
\end{align}

The problem can then be solved by a number of transformations, first
letting $B\left(r\right)=r^2 l\left(r\right)$ giving
\begin{equation}
-\frac{6\lambda^2}{\alpha_c\alpha^2}=r^6\left(8l'^2-r^2l''^2+2rl'\left(4l''+rl'''\right)\right).
\end{equation}

Then, consider $l'\left(r\right)=y\left(r\right)$ so that
\begin{equation}
-\frac{6\lambda^2}{\alpha_c\alpha^2}=r^6\left(8y^2-r^2y'^2+2ry\left(4y'+ry''\right)\right)
\end{equation}
reduces the overall order of the problem, and taking
$y\left(r\right)=r^{-3}h\left(r\right)$ gives a second-order
differential equation
\begin{equation}
-\frac{6\lambda^2}{\alpha_c\alpha^2}=-h^2-r^2h'^2+2rh\left(h'+rh''\right).
\end{equation}

As in Ref.\cite{0p01}, we consider an exponential transformation of
the form  $r=e^t$, which results in
\begin{equation}
-\frac{6\lambda^2}{\alpha_c\alpha^2}=-h^2-\dot{h}^2+2h\ddot{h},
\end{equation}
where dots denote derivatives with respect to $t$.
\newline

Following $h\left(t\right)=\left(v\left(t\right)\right)^2$, the equation
\begin{equation}
\frac{6\lambda^2}{\alpha_c\alpha^2}=v^3\left(v-4\ddot{v}\right),
\end{equation}
can be solved, where the first integral turns out to be
\begin{equation}
\frac{v^2}{2}+\frac{1}{2v^2}\frac{6\lambda^2}{\alpha_c\alpha^2}-2\left(\dot{v}\right)^2=c_1.
\label{1st_int}
\end{equation}
This admits two separate solutions given by
\begin{equation}
\!\!\!v\left(t\right)=\sqrt{-\frac{6\lambda^2}{\alpha_c\alpha^2}e^{-\left(t+c_2\right)}
+\frac{e^{t+c_2}}{4} - 2c_1 +4 c_1^2e^{-\left(t+c_2\right)}},
\label{sol1}
\end{equation}
and
\begin{equation}
\!\!\!v\left(t\right)=\sqrt{-\frac{6\lambda^2}{\alpha_c\alpha^2}e^{\left(t+c_2\right)}
+\frac{e^{-(t+c_2)}}{4} - 2c_1 +4 c_1^2e^{\left(t+c_2\right)}}
\label{sol2}.
\end{equation}

This means that when all the transformations are taken in reverse
and the solutions are represented with the coordinate $r$, the
solutions for $B\left(r\right)$ turn out to be
\begin{equation}
B\left(r\right)=\frac{2\lambda^2}{\alpha_c\alpha^2}\frac{e^{-c_2}}{r}-
\frac{e^{c_2}}{4}r +c_1 -\frac{4c_1^2e^{-c_2}}{3r} + c_3r^2,
\label{solb1}
\end{equation}
and
\begin{equation}
B\left(r\right)=\frac{6\lambda^2}{\alpha_c\alpha^2}e^{c_2}r-
\frac{e^{-c_2}}{12r} +c_1 -4c_1^2e^{c_2}r + c_3r^2,
\label{solb2}
\end{equation}
respectively, where $c_1$, $c_2$, and $c_3$ are constants. The
solution in Eq.(\ref{solb1}) can be represented in a form similar to
Eq.(\ref{lemos_metr}) as
\begin{equation}
B(r) = \frac{u}{r} + \sqrt{\frac{3\beta\gamma}{4}} + \frac{\gamma
r}{4} + k^2 r^2, \label{solb1f}
\end{equation}
where
\begin{equation}
u = \beta - \frac{2\lambda^2}{\alpha_{c}\alpha^2\gamma},
\end{equation}
and $k^2 = c_{3}$, $\gamma = -e^{c_2}$, and $c_1 =
\sqrt{\frac{3\beta\gamma}{4}}$. Similarly, Eq.(\ref{solb2}) takes the
form
\begin{equation}
B(r) = \frac{\beta}{r} + \sqrt{\frac{3\beta\gamma}{4}} +
\frac{\gamma r}{4} - \frac{\lambda^2 r}{2\alpha_{c}\alpha^2\beta} +
k^2 r^2, \label{solb2f}
\end{equation}
with $k^2 = c_3$, $c_1 = \sqrt{\frac{3\beta\gamma}{4}}$, and
$e^{c_2} = -\frac{1}{12\beta}$.

On comparison with the general relativity case in Eq.(\ref{lem_rel}),
the constants in Eqs.(\ref{solb1f}) and (\ref{solb2f}) regain their
regular meanings except for the $\gamma$ factor, which is a measure
of deviation from general relativity in conformal gravity. Also, in
this representation, it is easy to see that the two solutions are
a charged generalization of the neutral metric since setting $\lambda$
to zero retrieves Eq.(\ref{neut_metric}) in both cases.
\newline

In the first case given by Eq.(\ref{solb1f}), the coefficient $u$
retains its dependance on $\beta$  but adds charge by introducing a
length scale $1/\alpha_c$ which is the coupling constant of
conformal gravity as given in the action in Eq.(\ref{weyl_action}).
Meanwhile, the same linear term arises as in the neutral case.
However, the $1/r$ dependence of the charge term in $B(r)$ is
unexpected and shows a divergence between general relativity where
charge produces a $1/r^2$ term and conformal gravity where charge
(electromagnetism) and mass (gravitation) give rise to same
contribution to the exterior geometry. This will then make the task
of separating the effects on the motion of a test particle from the
two contributions less straightforward than in the case of general
relativity. We note that in both the second-order theory and the
fourth-order theory the only nonvanishing stress-energy tensor
component has the same form, namely, $\udt{T}{r}{r}\propto1/r^4$,
meaning that the different $1/r$ dependence of the charge term in
$B(r)$ emerges out of the geometric part of the fourth-order theory.
\newline

The same phenomenon was observed in the spherical case
\cite{riegert,furthersolutions1}, and so, since it also emerges here
in the cylindrical topology, it may suggest that this could be a
general feature of charged solutions in conformal gravity. This
would imply, as noted in Ref.\cite{furthersolutions1}, that
electromagnetism and gravitation may share a connection and some
underlying similarity.
\newline

The second solution given by Eq.(\ref{solb2f}) also possesses the same
 $\gamma r$ term as in the neutral metric, but now the charge
contribution to the external geometry features as an unexpected
additional linear term in the potential. This has the unrealistic
consequence that the effect from a charge distribution at the source
on the motion of a point charge increases with distance. On the
other hand, it may shed some light about the origin of linear terms
found practically in all known solutions of conformal Weyl gravity.
It is still unclear whether the $\gamma r$ terms in
Eqs.(\ref{mkmetric}), (\ref{neut_metric}), and now in Eqs.(\ref{solb1f}) and
(\ref{solb2f}) have a universal origin (like the cosmological term
$kr^2$) or are system dependent (like the mass term $\beta/r)$. From
Eq.(\ref{solb2f}) it seems that at least part of this is clearly system
dependent.

Finally, we compare our solutions in Eqs.(\ref{solb1f}) and
(\ref{solb2f}) with the earlier charged solutions in spherical
geometry obtained in Ref. \cite{riegert}, given by
\begin{align}
ds^2 & =-(ar^2 + br + c + d/r)dt^2 +\nonumber\\
 &(ar^2 + br + c + d/r)^{-1} dr^2 +r^2\,d\Omega_{2}^2,
\end{align}
where
\begin{equation}
3bd - c^2 + 1 + \frac{3\lambda^2}{2\alpha^2\alpha_{c}}=0.
\end{equation}
The above solution can be generalized to
\begin{align}
ds^2 & =-(ar^2 + br + c + d/r)dt^2 +\nonumber\\
 &(ar^2 + br + c + d/r)^{-1} dr^2 +r^2\,d\Omega_{2,K}^2,\label{general}
\end{align}
with
\begin{equation}
3bd - c^2 + K^2 +
\frac{3\lambda^2}{2\alpha^2\alpha_{c}}=0,\label{condition}
\end{equation}
where
\begin{equation}
d\Omega_{2,k}^2 = \frac{d\rho^2}{1 - K\rho^2} + \rho^2 d\phi^2
\end{equation}
represents the metric on a unit 2 sphere ($K = 1$), a unit
hyperbolic manifold ($K = -1$), or a 2 torus ($K = 0$). Comparing
Eqs.(\ref{solb1f}) and (\ref{solb2f}) with Eq.(\ref{general}) and putting
$K=0$, we found that both solutions satisfy Eq.(\ref{condition}).

\section{III. Temperature}
The most immediate path for studying the quantum nature of black
holes is through a consideration of their thermodynamical
properties. The horizon temperature, $T_h$, is the natural place to
start along this line of thought. This is defined in terms of the
surface gravity, $\kappa$, by the relation
$T_h=\frac{\kappa}{2\pi}$, which in turn is given in terms of the
Killing vector fields, $\chi^{\nu}$, by \cite{p12}
\begin{equation}
\kappa^2=-\frac{1}{2}\left(\nabla^{\mu}\chi^{\nu}\right)\left(\nabla_{\text{[}\mu}\chi_{\nu\text{]}}\right).
\end{equation}
The Killing vectors will be calculated at the horizon radius. For the first solution from Eq.(\ref{sol1}) the horizon radius is given by
\begin{widetext}
\begin{align}
&r_{1_h}=\frac{1}{12\alpha^2\beta\alpha_c k^2}\Bigg[2\lambda^2-\alpha^2\beta\alpha_c\gamma+\bigg(\alpha^4\beta^2\alpha_c^2\left(\gamma^2-24k^2\sqrt{3\beta\gamma}\right)\nonumber\\
&-4\alpha^2\beta\alpha_c\gamma\lambda^2+4\lambda^4\bigg)/\Bigg[6\alpha^4\beta^2\alpha_c^2\left(\gamma^2-12k^2\sqrt{3\beta\gamma}\right)\lambda^2-\alpha^6\beta^3\alpha_c^3\left(\gamma^3-36\gamma k^2\sqrt{3\beta\gamma}+864\beta k^4\right)\nonumber\\
&-12\alpha^2\beta\alpha_c\gamma\lambda^4+4\bar{\Sigma}\Bigg]^{1/3}+\Bigg[-\alpha^6\beta^3\alpha_c^3\left(\gamma^3-36\gamma k^2\sqrt{3\beta\gamma}+864\beta k^2\right)+6\alpha^4\beta^2\alpha_c^2\left(\gamma^2-12k^2\sqrt{3\beta\gamma}\right)\lambda^2\nonumber\\
&-12\alpha^2\beta\alpha_c\gamma\lambda^4+44\bar{\Sigma}\Bigg]^{1/3}\Bigg],
\end{align}
\end{widetext}
where
\begin{widetext}
\begin{align}
&\bar{\Sigma}=2\lambda^2+3\Bigg[3\alpha^6\beta^4\alpha_c^3k^4\Bigg(\alpha^6\beta^3\alpha_c^3\left(\gamma^3-48\gamma k^2\sqrt{3\beta\gamma}+1728\beta k^4\right)\nonumber\\
&-12\alpha^4\beta^2\alpha_c^2\left(\gamma^2-24 k^2\sqrt{3\beta\gamma}\right)\lambda^2+36\alpha^2\beta\alpha_c\gamma\lambda^4-32\lambda^6\Bigg)\Bigg]^{1/2},
\end{align}
\end{widetext}
while for the second given in Eq.(\ref{sol2}) the horizon radius is given by
\begin{widetext}
\begin{align}
&r_{2_h}=-\frac{\gamma}{12k^2}+\frac{1}{12k^2}\left(\gamma^2-24k^2\sqrt{3\beta\gamma}\right)\nonumber\\
&\Bigg[-\gamma^3+36k^2\sqrt{3\beta\gamma^3}-864k^4 u+12\sqrt{3k^4\left(-3\beta\left(\gamma^3-32k^2\sqrt{3\gamma^3\beta}\right)+4u\left(\gamma^3-36k^2\sqrt{3\gamma^3\beta}+432k^4u\right)\right)}\Bigg]^{-1/3}\nonumber\\
&+\frac{1}{12\sqrt[3]{2}k^2}\Bigg[-2\gamma^3+72k^2\sqrt{3\beta\gamma^3}-1728k^4u+\sqrt{-4\left(\gamma^2-24k^2\sqrt{3\beta\gamma}\right)^3+4\left(\gamma^3-36k^2\sqrt{3\beta\gamma^3}+864k^4 u\right)^2}\Bigg]^{-1/3}.
\end{align}
\end{widetext}
For either case of $r_{i_h}$ and with the metric in Eq.(\ref{con_cyl_met}), the following surface gravity is found:
\begin{equation}
\kappa=\frac{\gamma}{8}-\frac{\lambda^2}{4\alpha_c\alpha^2\beta}-\frac{\beta}{2r_{i_h}^2}+r_{i_h} k^2.
\end{equation}
Finally, this results in a horizon temperature
\begin{equation}
T_h=\frac{1}{2\pi}\left(\frac{\gamma}{8}-\frac{\lambda^2}{4\alpha_c\alpha^2\beta}-\frac{\beta}{2r_{i_h}^2+r_{i_h} k^2}\right),
\end{equation}
which for $\gamma=0$, reduces to the general relativistic result
\cite{p13}. Furthermore, for a vanishing charge density, we obtain the
expected result already found in Ref.\cite{0p01}.

\section{IV. Conclusion}
In this paper, we studied static charged cylindrical solutions in
conformal gravity, and we found that, unlike the second-order
Einstein's theory, there are two independent metric tensors which can
be used to describe the external geometry.

The field equations used for conformal gravity do not feature a
cosmological constant. However, both solutions contain the same
cosmological term which occurs in the general relativistic solution
(\ref{lem_rel}).

The linear $\gamma r$ term arises as in the neutral case
(\ref{neut_metric}) and the spherically symmetric solution in
Eq.(\ref{mkmetric}), where it was used \cite{conformal2} to
explain the flat rotational curves of galaxies and other large
matter distributions.
\newline

Moreover, we found that, unlike general relativity where the
modification from the charge to the static black string solution
(\ref{lemos_metr}) is in the form of a term which behaves like
$1/r^2$, in the fourth-order case, the modification to the neutral
solution (\ref{neut_metric}) is different. In one of the solutions,
the modification behaves like $1/r$, i.e., like a Newtonian term, as
in the spherically symmetric solution (\ref{mkcmetric}) derived
earlier by Mannheim and Kazanas, while in the second solution, the
modification is in the form of an additional linear term in the
metric.
\newline

The solutions found in this paper will not be applicable to most
astrophysical sources since, for the most part, they are organized
with spherical symmetry, but there are a number of other sources
which may be amenable to this background metric description. Another
place where cylindrical symmetry may arise is on the very small
scale since exotic forms of collapsing matter fields may take place
here. However, this type of collapse detection would, in all
likelihood, be in the form of Hawking radiation, which is one possible
avenue of future development. Moreover, there may also be stringlike
applications in a number of other theories such as with the use of
the AdS/CFT correspondence duality.

\section{Acknowledgments}
J. L. S. wishes to thank the Physics Department at the University of
Malta for hospitality during the completion of this work.
The authors acknowledge helpful suggestions from the referee
who improved the final version of this paper.

\end{document}